\documentstyle[pra,aps,twocolumn,psfig,floats]{revtex}
\begin{document}

\title{Optical implementation of continuous-variable quantum cloning machines}

\author{Jarom\'{\i}r Fiur\'{a}\v{s}ek$^{1,2}$}

\address{ {$^1$School~of~Physics~and~Astronomy,~University~of~St~%
Andrews,~North~Haugh,~St~Andrews,~Fife,~KY16~9SS,~Scotland }\\
$^2$Department of Optics, Palack\'{y} University, 17. listopadu 50,
772 07 Olomouc, Czech Republic}

\date{\today}

\maketitle

\begin{abstract}
 We propose an optical implementation of the Gaussian
continuous-variable quantum cloning machines. We construct a
symmetric $N\rightarrow M$ cloner which
optimally clones coherent states and we also provide an explicit
design of an asymmetric $1\rightarrow 2$ cloning machine. All proposed cloning devices
can be built from just a single non-degenerate optical parametric amplifier 
and several beam splitters.
\end{abstract}


Quantum information theory was originally developed in the
context of the discrete quantum variables, with the central notion
of a single qubit as a unit of quantum information. Recently,
however, various concepts of the quantum-information processing
have been extended to the domain of continuous quantum variables.
The established continuous-variable protocols
comprise the quantum state teleportation
\cite{BraunsteinT}, quantum error
correction \cite{BraunsteinEC}, quantum cryptography
\cite{Ralph99}, quantum computation
\cite{Lloyd99}, entanglement purification
\cite{Duan00},  quantum dense coding
\cite{BraunsteinDC}, and, finally, quantum cloning
\cite{Cerf00}.

In this paper we propose a simple optical implementation of the Gaussian
continuous-variable cloning machine introduced by Cerf. {\em et al.} \cite{Cerf00}.
It is well known that  the exact cloning of an unknown quantum
state is impossible \cite{Wootters82}. The cloning machine thus produces
approximate copies of the state subject to cloning.
We first  design an asymmetric $1\rightarrow 2$ cloning machine which produces
two copies of coherent state with different fidelities. Then we show how to build a symmetric 
$N\rightarrow M$ cloner which prepares $M$ identical approximate clones of $N$ available copies and optimally clones coherent states.
 It turns out that the cloning machines can
 be built from a single non-degenerate optical
parametric amplifier (NOPA, two-mode squeezer) and several beam splitters.

We note that the experimental realization of the Gaussian cloning
machine has been recently discussed in \cite{DAriano00} where
three NOPAs were used to build the machine. It is, however,
impossible to exactly construct the machine solely from two-mode
squeezeres. Therefore, the device proposed in \cite{DAriano00} is
only approximate, and the desired cloning transformation is
achieved only in the limit of infinite squeezing.  Such problems
are avoided if one employs beam splitters together with a two-mode
squeezer.

Let us begin with $1\rightarrow 2$ cloners.
Suppose that the mode $c$ initially contains an unknown quantum
state that we want to clone. At the output of the cloning machine,
two duplicates of the input state $|\psi\rangle_c$ emerge at the
modes $a$ and $c$. The implementation of cloning also requires an
ancilla mode $b$. At the input, modes $a$ and $b$  are in vacuum
states and the total three mode state can be written as
$|0\rangle_a|0\rangle_b|\psi_{\rm in}\rangle_c $. The unitary cloning
transformation performed by the cloning machine yields the output state
\cite{Cerf00,DAriano00}
\begin{equation}
 |\psi_{\rm out}\rangle =
 e^{-i(U+V)} e^{-i\chi Y}|0\rangle_a|0\rangle_b|\psi_{\rm in}\rangle_c.
\end{equation}
Here
\begin{eqnarray}
&U=i(c a^\dagger-c^\dagger a), \qquad  V=i(cb-c^\dagger b^\dagger),&
\nonumber \\
&Y=i(ab-a^\dagger b^\dagger),&
\end{eqnarray}
$a,b,c$ ($a^\dagger,b^\dagger,c^\dagger$) denote annihilation (creation) operators
of three different  modes of propagating optical field,
and $\chi$ is a parameter controlling the asymmetry of the two clones. If $\chi=(\ln 2)/2$
then the cloning is symmetric. It is thus convenient to define the new parameter
$\gamma=\chi-(\ln 2)/2$.

Notice that in Ref. \cite{Cerf00} the cloning transformation is
defined as $\exp[-i(U+V)]$ and the modes $a$ and $b$ enter the cloning machine
in an entangled two-mode squeezed state
$\exp(-i\chi Y)|0\rangle_a|0\rangle_b$. Here we assume that $a$ and $b$ are
initially in the vacuum state and we define the cloning transformation as
\begin{equation}
{\cal{C}}=\exp[-i(U+V)]\exp(-i\chi Y).
\end{equation}
The Hermitian operators $V$ and $Y$ are generators of two-mode squeeze transformations
which can be performed by NOPAs.
 $U$ is the generator of a mixing transformation, which can be implemented
  with the use of a beam splitter. It is easy to verify that $U,V$ and $Y$
  form a closed commutator algebra and they
thus generate a three parametric sub-group of the symplectic group $Sp(6,{\sf R})$.

For our purposes it is convenient to work in the Heisenberg
picture, where an arbitrary operator $X$ transforms according to
$X\rightarrow {\cal{C}}^\dagger X {\cal{C}}$.
After some algebra we obtain the cloning transformation in the Heisenberg picture,
\begin{eqnarray}
& a_{\rm out} = c_{\rm in}+\frac{\displaystyle e^\gamma}{\displaystyle\sqrt{2}}
(a_{\rm in}-b_{\rm in}^\dagger),&
\nonumber \\[2mm]
& b_{\rm out} = -\sqrt{2}\sinh\gamma\, a_{\rm in}^\dagger+
\sqrt{2} \cosh\gamma \, b_{\rm in}-c_{\rm in}^\dagger,&
 \nonumber \\[2mm]
& c_{\rm out} = c_{\rm in}-\frac{\displaystyle e^{-\gamma}}{\displaystyle \sqrt{2}}
(a_{\rm in}+b_{\rm in}^\dagger).&
\label{CLONING}
\end{eqnarray}
It was shown in \cite{Cerf00b} that the symmetric cloning machine corresponding to $\gamma=0$
in Eq. (\ref{CLONING}) is optimal for cloning of coherent states.

\begin{figure}[!t!]
\centerline{\psfig{figure=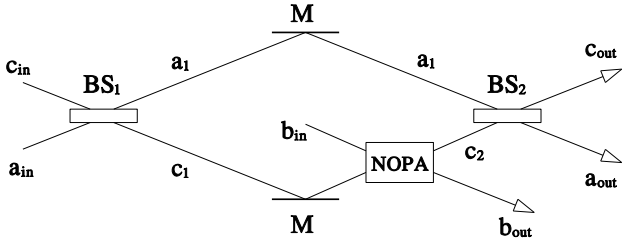,width=0.95\linewidth}}
\vspace*{2mm} \caption{Continuous-variable quantum cloning
machine. The device consists of two beam splitters BS$_1$ and
BS$_2$ and a non-degenerate parametric amplifiers (NOPA). M denote
auxiliary mirrors.}
\end{figure}

The cloning process (\ref{CLONING}) inevitably introduces  noise
into the two clones. To be specific, consider cloning of a coherent
state $|\xi\rangle$. It follows  from Eq. (\ref{CLONING}), that
the output single modes $a_{\rm out}$ and $c_{\rm out}$ are in
mixed states, namely  coherent states with superimposed thermal
noise with  mean number of chaotic photons
\begin{equation}
 \langle n_{{\rm ch}}\rangle_a= \frac{1}{2}e^{2\gamma},
\qquad
\langle n_{{\rm ch}}\rangle_c=\frac{1}{2}e^{-2\gamma}.
\end{equation}
A generic symplectic cloning transformation has the form
\begin{equation}
a_{\rm out}=c_{\rm in}+d+e^\dagger, \qquad
c_{\rm out}=c_{\rm in}+f+g^\dagger
\label{SYMPLECTIC}
\end{equation}
where $d,e,f,g$ commute with $c_{\rm in}^\dagger$.  We require
that the cloning (\ref{SYMPLECTIC}) does not prefer any direction
in the phase space so that clones
 of position and momentum states suffer from the same errors. This leads to the constraints
$[d,e^\dagger]=[f,g^\dagger]=0$. If this holds, then the Wigner functions of the
output modes $a_{\rm out}$ and $c_{\rm out}$ 
are convolutions of the Wigner function of $|\psi_{\rm in}\rangle_{c}$
with a phase independent Gaussian.
 One can show for generic cloning transformation (\ref{SYMPLECTIC})
that the product of noise photons is limited from below by
\begin{equation}
\langle n_{\rm ch} \rangle_a \langle n_{\rm ch} \rangle_c \geq \frac{1}{4}
\label{NOISELIMIT}
\end{equation}
and the transformation (\ref{CLONING}) achieves the lower bound $1/4$.
In this sense, the asymmetric cloner is optimal.

The Q-functions of the cloned coherent  states read
\begin{eqnarray}
Q_{a,c}(\alpha)=\frac{1}{(\langle n_{{\rm ch}}\rangle_{a,c}+1)\pi}
\exp\left(-\frac{|\alpha-\xi|^2}{\langle n_{{\rm ch}}
\rangle_{a,c}+1}\right)
\end{eqnarray}
and the  fidelities of the two clones of the coherent state $|\xi\rangle$
can be expressed as
\begin{eqnarray}
F_{a}=\langle \xi |\varrho_{a_{\rm out}}|\xi\rangle =\pi Q_a(\xi)=\frac{2}{e^{2\gamma}+2},
\nonumber \\
F_{c}=\langle \xi |\varrho_{c_{\rm out}}|\xi\rangle =\pi Q_c(\xi)=\frac{2}{e^{-2\gamma}+2}.
\end{eqnarray}
The fidelities are invariant for all coherent states. A symmetric
cloning machine has $\gamma=0$ and here both copies have the same
fidelity $F_a=F_c=2/3$ \cite{Cerf00}.

The explicit construction of the cloning machine can be based on
simple group-theoretical considerations. Since the cloning
transformation $I$ belongs to the three-parametric sub-group of
the group $Sp(6,{\sf R})$, we need three optical elements (beam
splitters and two-mode squeezers) to build a device performing
this transformation.  We construct the cloning
machine from two beam splitters and one squeezer because beam
splitters are much simpler and cheaper optical elements than
two-mode squeezers.

The cloning transformation $\cal{C}$ can be factorized as
\begin{equation}
{\cal{C}}= e^{-iwU} e^{-ivV} e^{-iuU} , \label{FACTORIZATION}
\end{equation}
where $u,v,w$ are unknown parameters that should be determined.
The factorization of a complicated symplectic transformation into
a sequence of simpler evolutions is a powerful technique which
allows us to get some physical insight into the nature of
$\cal{C}$. Recently, this approach has been applied to the analysis of
nonlinear optical couplers \cite{Fiurasek00}. Here we adopt the
same strategy to construct the cloning machine. A physical device
corresponding to the factorization (\ref{FACTORIZATION}) is
depicted in Fig. 1. The input modes $a_{\rm in}$ and $c_{\rm in}$
are mixed at the first beam splitter (BS$_1$) and $u$ is the
mixing angle of this beam splitter. Subsequently,  the modes $c_1$ and
$b_{\rm in}$ are squeezed in NOPA and $v$ is the squeezing parameter.
Finally, the modes $a_1$ and $c_2$ are re-combined at BS$_2$ whose
mixing angle is denoted by $w$.

 Remarkably, the cloning machine  depicted in Fig. 1
is essentially a Mach-Zhender interferometer with a NOPA placed in
one of its arms. The proposed scheme is also closely related to
the generic decomposition of linear optical circuits into sequence
of  $N$-port linear interferometer, $N$ single-mode squeezers, and
another $N$-mode interferometer put forward  by Braunstein
\cite{Braunstein99}. Our device has the added convenience of using
a single two-mode squeezer, which leads to the simplest possible structure.

In order to derive analytical formulas for the parameters $u,v,w$ as
 functions of $\gamma$ we consider the propagation of the output modes backward through
  the cloning machine.
When we transfer the output modes $a_{\rm out}$ and $c_{\rm out}$ back through  BS$_2$,
we get
\begin{eqnarray}
&a_1=\cos(w) a_{\rm out}-\sin(w) c_{\rm out},& \nonumber \\[1.5mm]
&c_2=\cos(w)c_{\rm out}+\sin(w) a_{\rm out}&
\label{A1C2}
\end{eqnarray}
On inserting (\ref{CLONING}) into (\ref{A1C2}), we have
\begin{eqnarray}
 a_1& =& \frac{1}{\sqrt{2}}\left(e^{\gamma} \cos w +e^{-\gamma}\sin
w\right) a_{\rm in}
       \nonumber \\
    &&+\frac{1}{\sqrt{2}}\left(e^{-\gamma}\sin w -e^{\gamma} \cos w\right) b_{\rm in}^\dagger
\nonumber \\   & &
 + (\cos w -\sin w) c_{\rm in},
 \label{A1C2EXP}
\end{eqnarray}
and a similar expression holds for $c_2$.
Before mixing at the second beam splitter, mode $a_1$ is not coupled to
$b_{\rm in}$. Therefore the coefficient in front of $b_{\rm in}^\dagger$ in Eq.
(\ref{A1C2EXP}) must be zero and we obtain
\begin{equation}
w= \arctan e^{2\gamma}, \qquad w\in\left[0,\frac{\pi}{2}\right].
\end{equation}

Next we employ a similar strategy to determine $v$. We transfer the modes 
$b_{\rm out}$
and $c_{\rm 2}$ back in front of the NOPA:
\begin{eqnarray}
&b_{\rm in}=\cosh(v) b_{\rm out}+\sinh(v) c_{\rm 2}^\dagger,& \nonumber \\[1.5mm]
& c_1=\cosh(v)c_{\rm 2}+\sinh(v) b_{\rm out}^\dagger.&
\end{eqnarray}
Since the mode $b_{\rm in}$ is not coupled to $c_{\rm in}$, we find
\begin{equation}
v= {\rm arg \tanh}\frac{\sqrt{1+e^{4\gamma}}}{1+e^{2\gamma}}.
\end{equation}

Finally, let us determine the mixing angle $u$. After mixing at the beam splitter, the
operator $a_1$ should read
\begin{equation}
a_1=a_{\rm in} \cos u + c_{\rm in} \sin u.
\end{equation}
Upon comparing this formula with (\ref{A1C2EXP}), we  find,
\begin{equation}
u=-\arctan(\sqrt{2}\sinh \gamma), \qquad   u\in\left[-\frac{\pi}{2},\frac{\pi}{2}\right].
\end{equation}

\begin{figure}[!t!]
\centerline{\psfig{figure=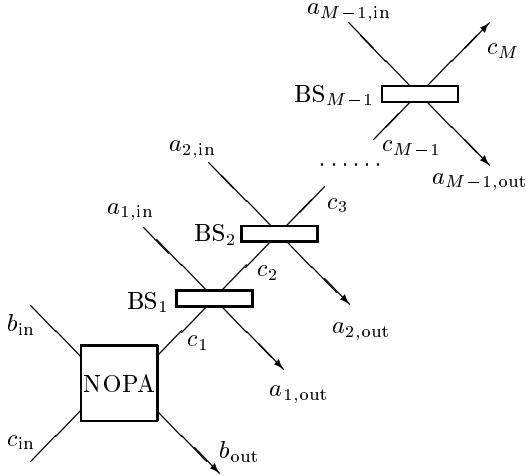,width=0.8\linewidth}}
\caption{Symmetric $1\rightarrow M$ quantum cloning machine. The device consists of a
non-degenerate parametric amplifier (NOPA) and $M-1$ beam splitters BS$_{j}$.
$M$ clones emerge in modes $a_{j,\rm out}$ and $c_M$.}
\end{figure}

The symmetric cloning machine with $\gamma=0$ has a particularly
simple structure. Since $u(\gamma=0)=0$, the machine can be built
as a sequence of a NOPA and a beam splitter. This result can be
readily generalized and we can design a symmetric cloning machine
which prepares $M$ identical approximate copies from a single
input state. As illustrated in Fig. 2, such a cloning machine is a
sequence of a NOPA and $M-1$ beam splitters with appropriately
adjusted transmittances and reflectances. The modes $a_{j, \rm in}$
and $b_{\rm in}$ are in the vacuum state. In the squeezer, the
amplitude of mode $c_{\rm in}$ is amplified by a factor of
$\sqrt{M}$, thus we have
\begin{eqnarray}
&c_{1}=\sqrt{M}\, c_{\rm in}-\sqrt{M-1} \, b_{\rm in}^\dagger,&
\nonumber \\
&b_{\rm out}=\sqrt{M} \, b_{\rm in}-\sqrt{M-1}\, c_{\rm in}^\dagger,&
\label{SYMNOPA}
\end{eqnarray}
After squeezing, the mode $c_1$ is mixed with $a_{1,\rm in}$ at the beam splitter BS$_1$.
The mode $a_{1,\rm out}$ contains the first approximate copy. The output $c_2$
is split at the BS$_2$ and so on, until we reach the last beam splitter BS$_{M-1}$.
The transformation performed by the jth beam splitter can be written as \cite{multiports}
\begin{eqnarray}
a_{j,\rm out}&=&\sqrt{\frac{1}{M-j+1}} \,c_{j}+\sqrt{\frac{M-j}{M-j+1}}\,a_{j,\rm in},
\nonumber \\[2mm]
c_{j+1}&=&\sqrt{\frac{M-j}{M-j+1}}\, c_j-\sqrt{\frac{1}{M-j+1}}\,a_{j,\rm in}.
\label{SYMOUT}
\end{eqnarray}
It is easy to verify that
\begin{eqnarray}
a_{j,\rm out}=c_{\rm in}-\sqrt{\frac{M-1}{M}} b_{\rm in}^\dagger+\sum_{k=1}^{j} \beta_{jk} a_{k,\rm in},
\nonumber \\
c_{M}=c_{\rm in}-\sqrt{\frac{M-1}{M}} b_{\rm in}^\dagger+\sum_{k=1}^{M-1}\beta_{Mk} a_{k,\rm in},
\end{eqnarray}
where the coefficients $\beta_{jk}$ can be determined from
(\ref{SYMOUT}). If $c_{\rm in}$ is prepared in the coherent state
$|\xi\rangle$ then the $Q$-function of all $M$ clones can be
expressed as
\begin{eqnarray}
Q(\alpha)=\frac{1}{\pi}\frac{M}{2M-1}
\exp\left(-\frac{M}{2M-1}|\alpha-\xi|^2\right)
\end{eqnarray}
and the cloning fidelity reads
$
F=\frac{M}{2M-1},
$
which is exactly the upper bound of the fidelity for  a $1\rightarrow
M$ cloner \cite{Cerf00b}. Thus we have proven that the machine
shown in Fig. 2 is the optimal one.

We are now in the position to address the most complicated case,
the optimal $N\rightarrow M$ cloner. The key observation is to
notice that the noise, which spoils the quality of each copy, stems
from the amplification in the NOPA. If $N$ copies of the coherent
state subject to cloning are available, then it would be optimal
to first collect all the signal into a single mode, then to amplify
this mode in a NOPA and, finally, to distribute the amplified signal
among $M$ output modes in the same manner as in Fig. 2. A scheme
of such an $N\rightarrow M$ cloner is given in Fig. 3. 
The $N$ available copies of the quantum state subj
ect 
to cloning are strored in modes $c_{1,\rm in},\ldots,c_{N,\rm in}$. 
The first
interferometer IF$_1$ contains a chain of $N-1$ beam splitters.
Let us denote by $d_{j+1}$ a mode propagating inside the IF$_1$
between $j$-th and $j+1$th beam splitters. The unitary
transformation performed by IF$_1$ can be written down as
\begin{eqnarray}
d_{j+1}=\sqrt{\frac{j}{j+1}} d_j+\sqrt{\frac{1}{j+1}}c_{j+1,\rm in},
\nonumber \\[1.5mm]
c_{j+1,\rm out}=\sqrt{\frac{1}{j+1}} d_j-\sqrt{\frac{j}{j+1}}c_{j+1,\rm in},
\end{eqnarray}
where $j=1,\ldots,N-1$, $d_1\equiv c_{1,\rm in}$ and $d_{N}\equiv c_{1,\rm out}$.
If the $N$ modes $c_{j,\rm in}$ are all in the same coherent state $|\xi\rangle$
then the $N-1$ output modes $c_{2,\rm out},\ldots,c_{N,\rm out}$ are in vacuum state
and all the signal is collected in mode $c_{1,\rm out}$ which is in the 
coherent state with amplitude $\sqrt{N} \xi$.

In the NOPA, the mode $c_{1,\rm out}$ is amplified by a factor of $\sqrt{M/N}$, and
we have
\begin{eqnarray}
e_{1}&=&\sqrt{\frac{M}{N}}\, c_{1,\rm out}-\sqrt{\frac{M-N}{N}} \, b_{\rm in}^\dagger,
\nonumber \\[1.5mm]
b_{\rm out}&=&\sqrt{\frac{M}{N}} \, b_{\rm in}-\sqrt{\frac{M-N}{N}}\, 
c_{1,\rm out}^\dagger,
\label{AMPMN}
\end{eqnarray}

\begin{figure}[!t!]
\centerline{\psfig{figure=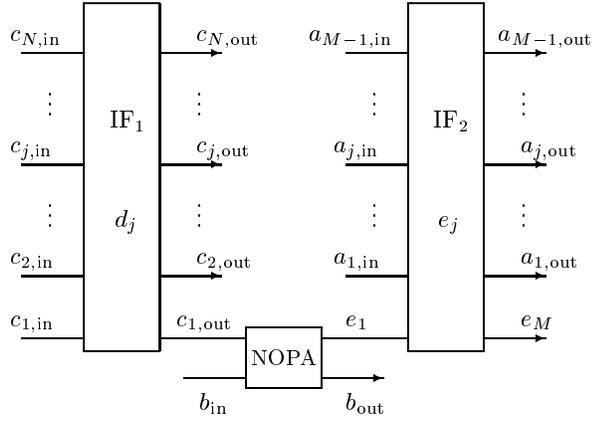,width=0.9\linewidth}}
\vspace*{2mm} \caption{Symmetric $N\rightarrow M$ quantum cloning
machine. The $N$-port interferometer IF$_1$ contains a sequence of
$N-1$ beam splitters. Similarly, the $M$-port interferometer
IF$_2$ is a chain of $M-1$ beam splitters. Mode $c_{1, \rm out}$
is amplified in non-degenerate parametric amplifier (NOPA).}
\end{figure}

Finally, the amplified signal is fed into a $M$-port interferometer
IF$_2$ where it is split into $M$ output modes, thereby preparing
$M$ approximate clones of the coherent state $|\xi\rangle$. The
interferometer IF$_2$ is exactly the sequence of $M-1$ beam
splitters
 shown in Fig. 3. If we use $e_j$ to denote a mode propagating between the $j$th
and the $j+1$th beam splitter inside IF$_2$ then the unitary
transformation performed by IF$_2$ is given by the formulas
(\ref{SYMOUT}) where we have to replace $c_j$ with $e_j$.
 To avoid possible confusion, we rewrite it explicitly,
\begin{eqnarray}
a_{j,\rm out}&=&\sqrt{\frac{1}{M-j+1}} \,e_{j}+\sqrt{\frac{M-j}{M-j+1}}\,a_{j,\rm in},
\nonumber \\[2mm]
e_{j+1}&=&\sqrt{\frac{M-j}{M-j+1}}\, e_j-\sqrt{\frac{1}{M-j+1}}\,a_{j,\rm in}.
\label{AE}
\end{eqnarray}
The $M$ approximate copies emerge in modes $a_{j,\rm out}$,
$j=1,\ldots,M-1$, and $e_{M}$. The amplification (\ref{AMPMN})
spoils the signal with $(M-N)/N$ chaotic photons which are equally
distributed among the $M$ copies. Thus each copy contains $\langle
n_{ch}\rangle =(M-N)/(MN)$ chaotic photons. Assuming an input
coherent state $|\xi\rangle$, all copies are prepared in
thermalized coherent state with the complex amplitude $\xi$ and
$\langle n_{ch}\rangle$ chaotic photons. The fidelity can be
obtained as
\begin{equation}
F=\frac{1}{\langle n_{\rm ch}\rangle+1}=\frac{MN}{MN+M-N},
\end{equation}
which is exactly the upper bound on fidelity of $N\rightarrow M$
cloner derived in \cite{Cerf00b}, hence the machine shown in Fig.
3 is an optimal $N\rightarrow M$ cloner of coherent states.

In summary, we have proposed an optical implementation of the
continuous-variable cloning machine. The asymmetric $1\rightarrow
2$ cloner is a Mach-Zhender interferometer with a NOPA in one of
its arms. The symmetric $N\rightarrow M$ cloner consists of
an $N$-port linear interferometer followed by a NOPA and an $M$-port
interferometer.

\vspace*{2mm}

I would like to thank Ulf Leonhardt for valuable comments.
This work was supported by Grant LN00A015 of the Czech Ministry of Education.
This paper is dedicated to the anniversary of 65th birthday 
of Prof. Jan Pe\v{r}ina.

\end{document}